\documentclass[10pt]{amsart}

\usepackage{verbatim}
\usepackage{eucal,url,amssymb,stmaryrd,
enumerate,amscd,
}

\usepackage[hypertex]{hyperref}      

\usepackage{amsfonts}
\usepackage{amsmath,amsthm,amssymb,amscd,enumerate,eucal,url,stmaryrd}

\numberwithin{equation}{section}

\newtheorem{thrm}{Theorem}[section]

\newtheorem{cor}[thrm]{Corollary}

\newtheorem{rmrk}[thrm]{Remark}

\begin{document}

\begin{abstract}
We show that  the heterotic supersymmetry (Killing spinor equations) and the anomaly cancellation  imply the
heterotic equations of motion in dimensions five, six, seven, eight if and only if the connection on the tangent bundle is an instanton. For heterotic compactifications in dimension six this reduces the choice of that connection to  the unique SU(3) instanton on a manifold with stable tangent bundle of degree zero.
\end{abstract}

\title[Heterotic supersymmetry, anomaly cancellation and equations of motion]
{Heterotic supersymmetry, anomaly cancellation\\ and equations of motion}
\date{\today}

\author{Stefan Ivanov}
\address[Ivanov]{University of Sofia ``St. Kl. Ohridski"\\
Faculty of Mathematics and Informatics\\
Blvd. James Bourchier 5\\
1164 Sofia, Bulgaria}
\address{and Department of Mathematics, University of Pennsylvania, Philadelphia, PA 19104-6395}
\email{ivanovsp@fmi.uni-sofia.bg}

\maketitle

\setcounter{tocdepth}{2} \tableofcontents

\section{Introduction. Field and Killing-spinor equations}

The bosonic fields of the ten-dimensional supergravity which arises
as low energy effective theory of the heterotic string are the
spacetime metric $g$, the NS three-form field strength $H$, the
dilaton $\phi$ and the gauge connection $A$ with curvature $F^A$.
The bosonic geometry considered in this paper is of the form
$R^{1,9-d}\times M^d$ where the bosonic fields are non-trivial only
on $M^d$, $d\leq 8$. One considers the two connections
$\nabla^{\pm}=\nabla^g \pm \frac12 H$,
where $\nabla^g$ is the Levi-Civita connection of the Riemannian
metric $g$. Both connections preserve the metric, $\nabla^{\pm}g=0$ and
have totally skew-symmetric torsion $T^{\pm}_{ijk}=g_{sk}(T^{\pm})^s_{ij}=\pm H_{ijk}$, respectively.

The bosonic part of the ten-dimensional supergravity action in the
string frame is \cite{Berg}
\begin{gather}\label{action}
S=\frac{1}{2k^2}\int
d^{10}x\sqrt{-g}e^{-2\phi}\Big[Scal^g+4(\nabla^g\phi)^2-\frac{1}{2}|H|^2
-\frac{\alpha'}4\Big(Tr |F^A|^2)-Tr |R|^2\Big)\Big],
\end{gather}
where $R$ is the curvature of a connection $\nabla$ on the tangent bundle and $F^A$ is the curvature of a connection $A$ on a vector bundle $E$.

The string frame field equations (the equations of motion induced
from the action \eqref{action}) of the heterotic string up to
two-loops \cite{HT} in sigma model perturbation theory are (we use
the notations in \cite{GPap})
\begin{gather}\nonumber
Ric^g_{ij}-\frac14H_{imn}H_j^{mn}+2\nabla^g_i\nabla^g_j\phi-\frac{\alpha'}4
\Big[(F^A)_{imns}(F^A)_j^{mns}-R_{imns}R_j^{mns}\Big]=0;\\\label{mot}
\nabla^g_i(e^{-2\phi}H^i_{jk})=0;\\\nonumber
\nabla^+_i(e^{-2\phi}(F^A)^i_j)=0,
\end{gather}
The field equation of the dilaton $\phi$ is implied from the first
two equations above.

A heterotic geometry will preserve supersymmetry if and only if, in
10 dimensions, there exists at least one Majorana-Weyl spinor
$\epsilon$ such that the supersymmetry variations of the fermionic
fields vanish, i.e. the following Killing-spinor equations hold
\cite{Str}
\begin{gather} \nonumber
\delta_{\lambda}=\nabla_m\epsilon = \left(\nabla_m^g
+\frac{1}{4}H_{mnp}\Gamma^{np} \right)\epsilon=\nabla^+\epsilon=0;
\\\label{sup1} \delta_{\Psi}=\left(\Gamma^m\partial_m\phi
-\frac{1}{12}H_{mnp}\Gamma^{mnp} \right)\cdot \epsilon=(d\phi-\frac12H)\cdot\epsilon=0; \\
\nonumber
\delta_{\xi}=F^A_{mn}\Gamma^{mn}\epsilon=F^A\cdot\epsilon=0,
\end{gather}
where  $\lambda, \Psi, \xi$ are the gravitino, the dilatino and the
gaugino  fields, respectively and $\cdot$ means Clifford action of
forms on spinors.

The instanton equation, the last equation in \eqref{sup1} means that the curvature 2-form $F^A$ is contained in the Lie algebra of the Lie group which is the stabilizer of the spinor $\epsilon$. It is known that in dimension 5,6,7 and 8 the stabilizer is the group $SU(2),SU(3),G_2$ and $Spin(7)$, respectively. An instanton (a solution to the last equation in \eqref{sup1}) in dimension 5,6,7 and 8 is a connection with curvature 2-from which is contained in the lie algebra $su(2),su(3),g_2$ and $spin(7)$, respectively \cite{CDev,Str,HS,DT,RC,GMW,DS}.

The Green-Schwarz anomaly cancellation mechanism requires that the
three-form Bianchi identity receives an $\alpha'$ correction of the
form
\begin{equation}\label{acgen}
dH=\frac{\alpha'}4
\Big(Tr(R\wedge R)-Tr(F^A\wedge F^A)\Big).
\end{equation}
A class of heterotic-string backgrounds for which the Bianchi
identity of the three-form $H$ receives a correction of type
\eqref{acgen} are those with (2,0) world-volume supersymmetry. Such
models were considered in \cite{HuW}. The target-space geometry of
(2,0)-supersymmetric sigma models has been extensively investigated
in \cite{HuW,Str,HP1}. Recently, there is revived interest in these
models \cite{GKMW,CCDLMZ,GMPW,GMW,GPap} as string backgrounds and in
connection to heterotic-string compactifications with fluxes
\cite{Car1,BBDG,BBE,BBDP,y1,y2,y3,y4}.

In writing \eqref{acgen} there is a subtlety to the choice of
connection $\nabla$ on $M^d$ since anomalies can be cancelled
independently of the choice \cite{Hull}. Different connections
correspond to different regularization schemes in the
two-dimensional worldsheet non-linear sigma model. Hence the
background fields given for the particular choice of $\nabla$ must
be related to those for a different choice by a field redefinition
\cite{Sen}. Connections on $M^d$ proposed to investigate the
anomaly cancellation  \eqref{acgen} are $\nabla^g$ \cite{Str,GMW},
$\nabla^+$ \cite{CCDLMZ}, $\nabla^-$ \cite{Berg,Car1,GPap,II},
Chern connection $\nabla^c$ when $d=6$ \cite{Str,y1,y2,y3,y4}.

It is known \cite{Bwit,GMPW} (\cite{GPap} for dimension $d=6$),
that the equations of motion of type I supergravity \eqref{mot}
with $R=0$ are automatically satisfied if one imposes, in addition
to the preserving supersymmetry equations \eqref{sup1}, the
three-form Bianchi identity \eqref{acgen} taken with respect to a flat
connection on $TM, R=0$.

According to no-go (vanishing) theorems  (a consequence of the
equations of motion \cite{FGW,Bwit}; a consequence of the
supersymmetry \cite{IP1,IP2} for SU($n$)-case and \cite{GMW} for the
general case) there are no compact solutions with non-zero flux and
non-constant dilaton satisfying simultaneously the supersymmetry
equations \eqref{sup1} and the three-form  Bianchi identity
\eqref{acgen} if one takes flat connection on $TM$, more precisely a
connection satisfying $Tr(R\wedge R)=0$.
Therefore, in the compact case one necessarily has to have a
non-zero term $Tr(R\wedge R)$. However, under the presence of a
non-zero curvature 4-form $Tr(R\wedge R)$ the solution of the
supersymmetry equations \eqref{sup1} and the anomaly cancellation
condition \eqref{acgen} obeys the second and the third equations of
motion but does not always satisfy the Einstein equation of motion
(the first equation in \eqref{mot}) \cite{GPap}. A quadratic expression for $R$
which is necessary and sufficient
condition in order that \eqref{sup1} and \eqref{acgen} imply
\eqref{mot} in dimension five, six, seven and eight are presented in
\cite{FIUV2,FIUV,FIUV1}. In particular, if
$R$ is an instanton the supersymmetry equations together with
the anomaly cancellation condition imply the equations of motion.

In  this note we show that the converse statement holds.
The main goal of the paper is to prove
\begin{thrm}\label{main}
The heterotic supersymmetry equations \eqref{sup1} together with the anomaly cancellation
\eqref{acgen} imply the heterotic equations of motion  \eqref{mot} on a manifold in dimensions
five, six, seven and eight if and only if the
connection on the tangent bundle in \eqref{acgen} is an  $SU(2), SU(3), G_2$ and $Spin(7)$
instanton in dimension five, six, seven and eight, respectively.
\end{thrm}
In the compact case in dimension six, it is shown in \cite[Theorem~1.1b]{FIUV} that the no-go theorems in \cite{IP1,IP2} force  the flux $H$ to vanish and the dilaton $\phi$ to be a constant for any compact solution to the heterotic supersymmetry \eqref{sup1} such that the (-)-connection on the tangent bundle is an $SU(3)$-instanton,  i.e. such a solution is a Calabi-Yau manifold. This result combined with Theorem~\ref{main} leads to
\begin{cor}\label{main1}
In dimension six, a compact solution  to the heterotic supersymmetry equations \eqref{sup1} satisfying anomaly cancellation \eqref{acgen} taken with respect to the (-)-connection imply the heterotic equations of motion \eqref{mot} if and only if  the flux $H$ is zero,  i.e. the solution is a Calabi-Yau manifold.
\end{cor}

\begin{rmrk}\label{main2}
Theorem~\ref{main} states that the heterotic equations of motion \eqref{mot} are consequences of the heterotic supersymmetry \eqref{sup1} and the anomaly cancellation \eqref{acgen} if and only if  the connection on the tangent bundle is of instanton type. On a compact solution to the gravitino and dilatino Killing spinor equations in dimension six, i.e. on a compact conformally balanced hermitian six-manifold with a holomorphic complex volume form \cite{Str} if there exists an $SU(3)$-instanton it is unique. Indeed, the non-K\"ahler version of the Donaldson-Uhlenbeck-Yau theorem \cite{Do,UY} established by Li-Yau \cite{LY} asserts via the Kobayashi-Hitchin correspondence that there exists an unique $SU(3)$-instanton (Yang-Mills connection) if and only if the holomorphic tangent bundle is stable of degree zero. Thus, Theorem~\ref{main} shows that the choice of the connection taken on the tangent bundle in \eqref{sup1} for compact supersymmetric heterotic solutions to \eqref{mot} in dimension six is fixed with the unique SU(3)-instanton.

This suggests that in order  to find compact heterotic supersymmetric solutions to the equations of motion \eqref{mot} in dimension six  one needs to start with a conformally balanced hermitian six manifold admitting holomorphic complex volume form with stable tangent bundle of degree zero and take the corresponding unique $SU(3)$-instanton in \eqref{acgen} and \eqref{action}.

Six dimensional compact supersymmetric solutions with non-zero flux $H$ and constant dilaton  of this kind are presented in \cite{FIUV}.
\end{rmrk}

In the context of perturbation theory the curvature $R^-$ of the (-)-connection is an one-loop-instanton due to the well known identity $R^+_{ijkl}-R^-_{klij}=\frac12dT_{ijkl}$, the first equation in \eqref{sup1} and \eqref{acgen} taken with respect to the (-)-connection. We thank the referee reminding this point to us. In this case, according to Theorem~\ref{main}, the supersymmetry \eqref{sup1} together with the anomaly cancellation \eqref{acgen} imply the heterotic equations of motion \eqref{mot} up to two loops. In fact the SU(3) case in dimension six has originally been dealt in \cite{GPap}. The $G_2$ case in dimension seven has been investigated in \cite[Section 6]{KO} when the anomaly cancellation has no zeroth  order terms in $\alpha'$. Compact up to two loops solutions in dimension six with non-zero flux $H$ and non-constant dilaton involving the (-)-connection are constructed in \cite{BBCG}.

If the anomaly cancellation has zeroth order term in $\alpha'$ (for example in  heterotic near horizons associated with $AdS_3$ investigated in the very recent paper \cite{GuP}) then $R^-$ is no longer one-loop instanton. In particular, in dimension six, Corollary~\ref{main1} and Remark~\ref{main2} is applicable suggesting a possible  lines for further investigations.

One can take the anomaly contribution  which appears at order $\alpha'$ as exact. Suppose that \eqref{acgen} is exact in the first order in $\alpha'$. Then, in dimension six  Corollary~\ref{main1} applies and  arguments  in Remark~\ref{main2} could be helpful  in further developments.

{\bf Conventions:} We choose a local orthonormal frame  $e_1,\dots,e_d$,  identifying it  with the dual
basis via the metric and write $e_{i_1 i_2\dots i_p}$ for the
monomial $e_{i_1} \wedge e_{i_2} \wedge \dots \wedge e_{i_p}$.

We  rise and lower the indices with the metric
and use the summation convention on repeated indices. For example,
$B_{ijk}C^{ijk}=B_i^{jk}C^i_{jk}=B_{ijk}C_{ijk}=
\sum_{ijk=1}^nB_{ijk}C_{ijk}.$

For a p-form $\beta$ we have the convention
$\beta=\frac1{p!}\beta_{i_1 i_2\dots i_p}e_{i_1 i_2\dots i_p}$.

The tensor norm  is denoted with $||.||^2$. For example
$||B||^2=B_{ijk}B^{ijk}=B_i^{jk}B^i_{jk}=B_{ijk}B_{ijk}.$

The curvature 2-forms $R_{ij}$ of a connection $\nabla$ are defined by $R_{ij}=[\nabla_i,\nabla_j]-\nabla_{[i,j]},
\quad R_{ijkl}=R^s_{ijk}g_{ls}.
$

The 4-form $Tr(R\wedge R)$ reads $ Tr(R\wedge
R)_{ijkl}=2\Big(R_{ijab}R_{klab}+R_{jkab}R_{ilab}+R_{kiab}R_{jlab}
\Big). $

The Hodge star operator on a $d$-dimensional manifold is denoted by $*_d$.
\section{Geometry of the heterotic supersymmetry}

Geometrically, the vanishing of the gravitino variation is
equivalent to the existence of a non-trivial real spinor parallel
with respect to the metric connection $\nabla^+$ with totally
skew-symmetric torsion $T=H$. The presence of $\nabla^+$-parallel
spinor leads to restriction of the holonomy group $Hol(\nabla^+)$ of
the torsion connection $\nabla^+$. Namely, $Hol(\nabla^+)$ has to be
contained in $SU(2), d=5$ \cite{FI,FI2,FIUV2}, $SU(3), d=6$
\cite{Str,IP1,IP2,GMW,GIP,CCDLMZ,BBDG,BBE}, the exceptional group
$G_2, d=7$ \cite{FI,GKMW,GMW,FI1}, the Lie group $Spin(7), d=8$
\cite{GKMW,I1,GMW}. A detailed analysis of the induced  geometries
is carried out in \cite{GMW} and all possible geometries (including
non compact stabilizers) are investigated in \cite{GLP,GPRS,GPR,P}.

A consequence of the gravitino and dilatino Killing spinor equations is  an
expression of the Ricci tensor $Ric^+_{mn}=R^+_{imnj}g^{ij}$ of the
(+)- connection, and therefore an expression of the Ricci tensor
$Ric^g$ of the Levi-Civita connection, in terms of the suitable
trace of the torsion three-form $T=H$ (the Lee form) and the
exterior derivative of the torsion form $dT=dH$ (see \cite{FI} for
dimensions 5 and 7, \cite{IP1} for dimension 6 (more precisely for any even dimension)
and \cite{I1} for dimension 8 as well as \cite{FIUV2,FIUV,FIUV1}).

We recall that the  Ricci tensors of $\nabla^g$ and $\nabla^+$ are connected by (see e.g. \cite{FI,FIUV1})
\begin{gather}\label{mo}
Ric^g_{mn}=\frac12(Ric^+_{mn}+Ric^+_{nm})+\frac14T_{mpq}T_n^{pq},\quad Ric^+_{mn}-Ric^+_{nm}=(\delta T)_{mn}=-(*_dd*_dT)_{mn}
\end{gather}

\subsection{Dimension five. Proof of Theorem~\ref{main} in $d=5$.}
The existence of $\nabla^+$-parallel spinor in dimension 5
determines an almost contact metric structure whose properties as
well as solutions to gravitino and dilatino Killing-spinor equations
are investigated in \cite{FI,FI2,FIUV2}.

We recall that an almost contact metric structure consists of an odd
dimensional manifold $M^{2k+1}$ equipped with a Riemannian metric
$g$, vector field $\xi$ of length one, its dual 1-form $\eta$ as
well as an endomorphism $\psi$ of the tangent bundle such that
$\psi(\xi)=0, \quad \psi^2=-id +\eta\otimes\xi, \quad
g(\psi.,\psi.)=g(.,.)-\eta\otimes\eta.
$ In local coordinates we also have
$\psi^i_j\xi^j=0, \quad \psi^i_s\psi^s_j=-\delta^i_j +\eta_j\xi^i, \quad
g_{st}\psi^s_i\psi^t_j=g_{ij}-\eta_i\eta_j.
$
The Reeb vector field $\xi$ is determined by the equations
$\eta(\xi)=\eta_s\xi^s=1,\quad (\xi\lrcorner d\eta)_i=d\eta_{si}\xi^s=0$,
where $\lrcorner$ denotes the interior multiplication.
The fundamental form $F$ is defined by $F(.,.)=g(.,\psi.), \quad F_{ij}=g_{is}\psi^s_j$
and the Nijenhuis tensor $N$   of an almost
contact metric structure is given by
$
 N=[\psi.,\psi.]+\psi^2[.,.] -\psi
[\psi.,.]-\psi[.,\psi.]  +d\eta\otimes\xi.
$

An almost contact metric structure is called normal if $N=0$; contact if $d\eta=2F$; quasi-Sasaki  if $N=dF=0$; Sasaki if $N=0, d\eta=2F$. The Reeb vector field $\xi$ is  Killing in the last two cases \cite{Bl}.

An almost contact metric structure admits a linear connection
$\nabla^+$ with torsion 3-form preserving the structure, i.e.
$\nabla^+ g=\nabla^+\xi=\nabla^+\psi=0$, if and only if the Nijenhuis
tensor is totally skew-symmetric, and the vector field $\xi$ is a
Killing vector field \cite{FI}. In fact, if the Nijenhuis tensor is totally skew-symmetric
then $\xi$ is a Killing vector field  exactly when (\cite[Proposition~3.1]{FI2},\cite{FIUV2})
\begin{equation}\label{simpl}
(\xi\lrcorner dF)_{ij}=dF_{sij}\xi^s=0 \Leftrightarrow (\xi\lrcorner N)_{ij}=N_{sij}\xi^s=0.
\end{equation}
In this case the torsion
connection is unique. The torsion $T$ of $\nabla^+$  is expressed by (\cite{FI,FI2,FIUV2})
$
T=\eta\wedge d\eta+d^{\psi}F+N,
$
where $d^{\psi}F=-dF(\psi.,\psi.,\psi), \quad (d^{\psi}F)_{ijk}=-dF_{str}\psi^s_i\psi^t_j\psi^r_k$.
In particular one has
$ d\eta_{ij}=(\xi\lrcorner T)_{ij}=T_{sij}\xi^s, \quad (\xi\lrcorner d\eta)_i=T_{sti}\xi^s\xi^t=0,\quad
d\eta(.,.)=d\eta(\psi.,\psi.), \quad d\eta_{ij}=d\eta_{st}\psi^s_i\psi^t_j$.

Since $\nabla^+\xi=0$ the restricted holonomy group $Hol(\nabla^+)$ of $\nabla^+$ contains  in $U(k)$ and
$Hol(\nabla^+)\subset SU(k)$ is equivalent to the following curvature  condition  found in \cite[Proposition 9.1]{FI}
\begin{equation}\label{su2}
R^+_{ijkl}F^{kl}=0 \Leftrightarrow R^+_{kijk}=Ric^+_{ij}=-\nabla^+_i\theta^5_j-\frac14\psi^s_jdT_{islm}F^{lm},
\quad \theta^5_i=\frac12\psi^s_iT_{skl}F^{kl}=\frac12dF_{ikl}F^{kl}.
\end{equation}
where $\theta^5$ is the Lee form defined in \cite{FI2}. Consequently, $\theta^5(\xi)=0$.

In  dimension five the Nijenhuis tensor is totally skew-symmetric exactly when it
vanishes \cite{CM}.  In this case $\xi$ is
a Killing vector field \cite{Bl}, the Lee form determines completely
the three form $dF$ due to \eqref{simpl}, $dF=\theta^5\wedge F$.
The dilatino equation
admits a solution if and only if  (\cite{FI2}, Proposition 5.5)
\begin{equation}\label{dilz}
2d\phi=\theta^5, \quad  *_{\mathbb H}d\eta=-d\eta,
\end{equation}
where  $*_{\mathbb H}$ denote the Hodge operator acting  in the
4-dimensional orthogonal complement $\mathbb H$ of the vector $\xi$,
$\mathbb H=Ker \eta$.  In particular, there is
no solution on any Sasaki 5-manifold.

The torsion (the NS three-form $H$) of a solution to gravitino and dilatino Killing spinor equations in dimension five is given by \cite{FI,FIUV2}
\begin{equation}\label{tsol5}
H=T=\eta\wedge d\eta +2d^{\psi}\phi\wedge F.
\end{equation}

An equivalent formulation is presented in \cite{FIUV2}.
The gravitino Killing spinor equation defines a reduction of the
structure group $SO(5)$ to $SU(2)$ which is described in terms of
forms by Conti and Salamon in \cite{ConS} (see also \cite{GGMPR}) as
follows: an $SU(2)$-structure on 5-dimensional manifold $M^5$ is
$(\eta,\omega_1,\omega_2,\omega_3)$, where $\eta$ is a $1$-form and $\omega_1,\omega_2,\omega_3$
are $2$-forms on $M$ satisfying $
  \omega_q\wedge \omega_r=\delta_{qr}v, \quad q,r=1,2,3, \quad
  v\wedge\eta\not=0,$
for some $4$-form $v$, and $X\lrcorner \omega_1=Y\lrcorner
\omega_2\Rightarrow \omega_3(X,Y)\ge 0.$

The gravitino and dilatino  Killing-spinor equations have a solution
exactly when there exists an $SU(2)$-structure
($\eta,\omega_1,\omega_2,\omega_3$) satisfying \cite{FIUV2}
$
d\omega_p=\theta^5\wedge \omega_p, \quad \theta^5(\xi)=0, \quad \theta^5=2d\phi, \quad
*_{\mathbb H}d\eta = - d\eta.
$
This means that the 'conformal' structure $\bar\eta=\eta,\bar\omega_p=e^{-2\phi}\omega_p$ is quasi Sasaki with $*_{\mathbb H}d\eta = - d\eta$.

In addition to these equations, the vanishing of the gaugino
variation requires the 2-form $F^A$ to be of instanton type
\cite{CDev,Str,HS,DT,RC,GMW,DS}. In dimension five, an $SU(2)$-instanton is
a connection $A$ with curvature two form  $F^A\in su(2)$. The $SU(2)$-instanton condition reads
\begin{equation}\label{inst5}
\begin{aligned}
(\xi\lrcorner F^A)_n=\xi^sF^A_{sn}=0, \quad
F^A(e_i,\psi e_i)=F^A_{st}F^{st}=0,\quad \psi^s_m\psi^t_nF^A_{mn}-F^A_{st}=0.
\end{aligned}
\end{equation}
\subsubsection{Theorem~\ref{main} in dimension 5}
\begin{proof}
We have to investigate only the Einstein equation of motion in dimension 5. First we observe that
$dd^{\psi}\phi(\xi,X)=-\xi\psi X\phi+\psi[\xi,X]\phi=0$, where we applied to the dilaton $\phi$ the identity  $0=(\mathbb L_{\xi}\psi)X =
[\xi,\psi X]-\psi[\xi,X]$, $\mathbb L$ is the Lie derivative, valid on any normal almost contact manifold \cite{Bl}, and  use $\xi(\phi)=0$. Then we  calculate from \eqref{tsol5} that $\psi^s_jdT_{islm}F^{lm}=-4d\eta_{si}d\eta_{sj}+\Big[(2dd^{\psi}\phi)_{st}F^{st}-8||d\phi||^2\Big]g_{ij}$ which implies
$\psi^s_jdT_{islm}F^{lm}=\psi^s_idT_{jslm}F^{lm}$.
Use the latter identity, substitute
\eqref{dilz} into the second equation of \eqref{su2} and the
obtained equality insert into \eqref{mo} using $2\nabla^g=2\nabla^+-T$
to get \cite{FIUV2}
\begin{gather}\label{mo5}
Ric^g_{ij}=-2\nabla^g_id\phi_j-\frac14\psi^s_jdT_{islm}F^{lm}+\frac14T_{mpq}T_n^{pq}.
\end{gather}
Substitute \eqref{acgen}  into \eqref{mo5}, use  \eqref{inst5} and
compare the result with the first equation in \eqref{mot} to
conclude that the supersymmetry equations \eqref{sup1} together
with the anomaly cancellation \eqref{acgen} imply the first
equation in \eqref{mot} if and only if the next equality holds
\cite{FIUV2}
\begin{equation}\label{supmot}
\begin{aligned}
R_{mstr}R_n^{str}=\frac1{2}\Big[R_{msij}R_{trij}+R_{mtij}R_{rsij}+R_{mrij}R_{stij}\Big]F^{tr}\psi^s_n.
\end{aligned}
\end{equation}
Multiplying \eqref{supmot} with $\xi^m\xi^n$ we obtain
$||\xi^mR_{mijk}||^2=0$. Hence, $\xi\lrcorner R=0$ which implies the first equation in
\eqref{inst5}. Thus, the curvature 2-form $R_{ij}$ is defined on
$\mathbb H$. The restriction of $\psi$ on $\mathbb H,\quad
\psi|_{\mathbb H}$ is an almost complex structure on $\mathbb H$.
The curvature two-form $R_{ij}$  decomposes into two
orthogonal parts $R'$ and $R''$ under the action of $\psi$ as follows
\begin{equation}\label{decomsu2}
\begin{aligned}
R'_{ij}=\frac12(R_{ij}+\psi^s_i\psi^t_jR_{st}),\quad R''_{ij}=\frac12(R_{ij}-\psi^s_i\psi^t_jR_{st}),\quad
\psi^s_i\psi^t_jR'_{st}=R'_{ij}, \quad \psi^s_i\psi^t_jR''_{st}=- R''_{ij}.
\end{aligned}
\end{equation}
An application of  \eqref{decomsu2} to \eqref{supmot} yields
\begin{equation*}\label{su2fin}
\begin{aligned}
2(||R'||^2+||R''||^2)=2R_{mstr}R_m^{str}=-||R_{mstr}F^{ms}||^2+2||R'||^2-2||R''||^2.
\end{aligned}
\end{equation*}
Consequently, $||R_{mstr}F^{ms}||^2+4||R''||^2=0$ which is equivalent to the  second and the third  equalities in \eqref{inst5}. Hence, $R$ is an $SU(2)$-instanton.
\end{proof}

\subsection{Dimension six. Proof of Theorem~\ref{main} in $d=6$.}
The necessary and sufficient condition for the existence of
solutions to the first two equations in \eqref{sup1} in an even dimension were derived
by Strominger \cite{Str} and investigated by many authors since then. Solutions are complex
conformally balanced manifold with non-vanishing holomorphic
volume form  satisfying an additional condition.

In dimension six any solution to the gravitino Killing spinor
equation reduces the holonomy group $Hol(\nabla^+)\subset SU(3)$. This defines  an almost hermitian structure $(g,J)$ with non-vanishing complex volume form  \cite{Str} which is preserved by the torsion connection.  We adopt for the  K\"ahler form $ \Omega_{ij}=g_{is}J^s_j.$ The Lee form $\theta^6$ is defined by $
\theta^6_i=-(*_6d*_6\Omega)_sJ^s_i=\frac12d\Omega_{ist}\Omega^{st}.
$

An almost hermitian structure admits a (unique) linear connection
$\nabla^+$ with torsion 3-form preserving the structure, i.e.
$\nabla^+ g=\nabla^+J=0$, if and only if the Nijenhuis
tensor is totally skew-symmetric \cite{FI}).

In addition, the dilatino equation forces the almost complex
structure to be integrable and the Lee form to be exact determined
by the dilaton. The torsion (the NS three-form $H$) is given by \cite{Str}
\begin{equation}\label{tsol6}
H_{ijk}=T_{ijk}=-J^s_iJ^t_jJ^r_kd\Omega_{str}, \qquad \theta^6_i=2d\phi_i=\frac12J^k_iT_{kst}\Omega^{st}.
\end{equation}

Since $\nabla^+ g=\nabla^+J=0$ the restricted holonomy group $Hol(\nabla^+)$ of $\nabla^+$ contains  in $U(k)$ and
$Hol(\nabla^+)\subset SU(k)$ is equivalent to the next curvature  condition  found in \cite[Proposition~3.1]{IP1}
\begin{equation}\label{su3}
R^+_{ijkl}\Omega^{kl}=0 \Leftrightarrow Ric^+_{ij}=-\nabla^+_i\theta^6_j-\frac14J^s_jdT_{islm}\Omega^{lm}.
\end{equation}

In addition to these equations, the vanishing of the gaugino
variation requires the 2-form $F^A$ to be of instanton type. In dimension six, an $SU(3)$-instanton (or a hermitian-Yang-Mils connection) is
a connection $A$ with curvature two form  $F^A\in su(3)$. The $SU(3)$-instanton condition is
\begin{equation}\label{inst6}
\begin{aligned}
F^A(e_i,Je_i)=F^A_{st}\Omega^{st}=0, \qquad
J^s_mJ^t_nF^A_{mn}-F^A_{st}=0.
\end{aligned}
\end{equation}

In complex coordinates the condition \eqref{inst6} reads
$F^A_{\mu\nu}=F^A_{\bar{\mu}\bar{\nu}}=0, \quad
F^A_{\mu\bar{\nu}}\Omega^{\mu\bar{\nu}}=0$ which is the well known Donaldson-Uhlenbeck-Yau instanton.

\subsubsection{Theorem~\ref{main} in dimension 6}

\begin{proof}
We need to investigate the Einstein equation of motion in dimension 6. Substitute the second equation of
\eqref{tsol6} into  \eqref{su3} and the
obtained equality insert into \eqref{mo} and use $2\nabla^g=\nabla^+-T$
to get \cite{IP1}
\begin{gather}\label{mo6}
Ric^g_{ij}=-2\nabla^g_id\phi_j-\frac14J^s_jdT_{islm}\Omega^{lm}+\frac14T_{mpq}T_n^{pq},
\end{gather}
where we used that on a complex manifold $dT=2\sqrt{-1}\partial\bar\partial \Omega$ is a (2,2)-form and therefore $J^s_jdT_{islm}\Omega^{lm}$ is symmetric in $i$ and $j$.

Substitute \eqref{acgen}  into \eqref{mo6}, use  \eqref{inst6} and
compare the result with the first equation in \eqref{mot} to
conclude that the supersymmetry equations \eqref{sup1} together
with the anomaly cancellation \eqref{acgen} imply the first
equation in \eqref{mot} if and only if the next equality holds
\cite{FIUV}
\begin{equation}\label{supmot6}
R_{mstr}R_n^{str}=\frac12\Big[R_{msij}R_{trij}+R_{mtij}R_{rsij}+R_{mrij}R_{stij}\Big]\Omega^{tr}J^s_n.
\end{equation}
The  two-form $R_{ij}$  decomposes into two
orthogonal parts $R'$ and $R''$ under the action of $J$ as follows
\begin{equation}\label{decomsu3}
\begin{aligned}
R'_{ij}=\frac12(R_{ij}+J^s_iJ^t_jR_{st}),\quad R''_{ij}=\frac12(R_{ij}-J^s_iJ^t_jR_{st}),\quad
J^s_iJ^t_jR'_{st}=R'_{ij}, \quad J^s_iJ^t_jR''_{st}=- R''_{ij}.
\end{aligned}
\end{equation}
We derive from \eqref{supmot6} and  \eqref{decomsu3} that
\begin{equation*}\label{su3fin}
\begin{aligned}
2(||R'||^2+||R''||^2)=2R_{mstr}R_m^{str}=-||R_{mstr}\Omega^{ms}||^2+2||R'||^2-2||R''||^2.
\end{aligned}
\end{equation*}
Hence, $|R_{mstr}\Omega^{ms}||^2+4||R''||^2=0$ which is precisely the $SU(3)$-instanton condition
\eqref{inst6} for $R$.
\end{proof}
\begin{rmrk}
Note that Theorem~\ref{main}, Corollary~\ref{main1} and Remark~\ref{main2} are valid for any even dimension.
\end{rmrk}

\subsection{Dimension seven. Proof of Theorem~\ref{main} in $d=7$.}
The existence of $\nabla^+$-parallel spinor in dimension 7
determines a $G_2$ structure whose properties as
well as solutions to gravitino and dilatino Killing-spinor equations
are investigated in \cite{FI,GKMW,FI1,GMPW,GMW,FIUV1}.

We briefly recall the notion of a $G_2$ structure.
Consider the three-form $\Theta$
on ${\mathbb R}^7$ given by
\begin{equation*}
  \Theta =e_{127} - e_{236} + e_{347}+ e_{567} - e_{146} - e_{245} +
  e_{135}.\label{11}
\end{equation*}
The subgroup of $GL(7,\mathbb{R})$ fixing $\Theta$ is the
 Lie group $G_2$ of dimension 14.    The 3-form
$\Theta$ corresponds to a real spinor  and therefore,
$G_2$ can be identified as the isotropy group of a non-trivial
real spinor.

The Hodge star operator supplies the 4-form $*_7\Theta$ given by
\begin{equation*}
  *_7\Theta =  e_{3456} + e_{1457} + e_{1256}+ e_{1234} + e_{2357} +
  e_{1367} - e_{2467}.\label{12}
\end{equation*}
We have the well known formula (see e.g. \cite{Br1,GMW,CI1,CI2})
\begin{equation}\label{g2-4-form}
*_7\Theta_{ijpq}*_7\Theta_{klpq} = 4\delta_{ik}\delta_{jl} -
  4\delta_{il}\delta_{jk} + 2*_7\Theta_{ijkl}.
\end{equation}
A $7$-dimensional Riemannian manifold $M$ is called a
$G_2$-manifold if its structure group reduces to the exceptional
Lie group $G_2$. The existence of a $G_2$-structure is equivalent
to the existence of a global non-degenerate three-form which can
be locally written as \eqref{11}.

If $\nabla^g\Theta=0$ then the Riemannian holonomy group is
contained in $G_2$.  It was shown by Gray~\cite{Gr} (see
also~\cite{FG,Br,Sal}) that this condition is equivalent to $d\Theta=d*\Theta=0$.
The Lee form $\theta^7$ is defined by \cite{Cabr}
$ \theta^7=-\frac{1}{3}*_7(*_7d\Theta\wedge\Theta) = \frac{1}{3}*_7(*_7
d*_7\Theta\wedge*_7\Theta). $

The precise conditions to have a solution to the gravitino Killing
spinor equation in dimension 7 were found in \cite{FI}. Namely,
there exists a non-trivial parallel spinor with respect to a
$G_2$-connection with torsion 3-form $T$ if and only if there
exists a $G_2$-structure $\Theta$ satisfying
$d*_7\Theta=\theta^7\wedge *_7\Theta$. In this case, the torsion
connection $\nabla^+$ is unique and the torsion 3-form $T$ is
given by $T=\frac{1}{6}(d\Theta,*_7\Theta)\,\Theta - *_7d\Theta
+*_7(\theta^7\wedge\Theta).$ Applying Theorem~4.8 in \cite{FI} and
the identity $*_7(\theta^7\wedge\Theta)=-(\theta^7\lrcorner
*_7\Theta)$ we can  write
\begin{equation}\label{g2liad}
\theta^7_s=-\frac{1}{18}((*_7d\Theta)_{ijk}*_7\Theta_{sijk}), \qquad
T_{ijk}*_7\Theta_{sijk}=-6\theta^7_s.
\end{equation}
The necessary conditions to have a solution to the system of
dilatino and gravitino Killing spinor equations were derived in
\cite{GKMW,FI,FI1}, and the sufficiency was proved in
\cite{FI,FI1}. The general  result \cite{FI,FI1} states
that there exists a non-trivial solution to both dilatino and
gravitino Killing spinor equations in dimension 7 if and only if
there exists a $G_2$-structure $\Theta$  satisfying the
equations $ d*_7\Theta=\theta^7\wedge *_7\Theta, \quad
d\Theta\wedge\Theta=0, \quad \theta^7=2d\phi, $ i.e. the conformal $G_2$-structure
$(\bar\Theta=e^{-\frac32\phi}\Theta,\bar g=e^{-\phi}g)$ obeys the
equations $d\bar{*}\bar\Theta=d\bar\Theta\wedge\bar\Theta=0$.

The
the flux $H$ of a solution to the gravitino and dilatino killing spinor equations is \cite{GKMW,FI,FI1}
\begin{equation}\label{tsol7}
H=T= -*_7d\Theta + 2*_7(d\phi\wedge\Theta).
\end{equation}
The Ricci tensor of the torsion connection was calculated in
\cite{FI} (see also \cite{FIUV1})
\begin{equation}\label{ric7}
Ric^+_{mn}=\frac1{12}dT_{mjkl}*_7\Theta_{njkl}+\frac16\nabla^+_mT_{jkl}*_7\Theta_{njkl}.
\end{equation}
Using the special expression of the torsion \eqref{tsol7} and
\eqref{g2liad}  the equation \eqref{ric7} takes
the form
\begin{equation}\label{ric77}
Ric^+_{mn}=\frac1{12}dT_{mjkl}*_7\Theta_{njkl}-2\nabla^+_md\phi_n=\frac1{12}dT_{mjkl}
*_7\Theta_{njkl}-2\nabla^g_md\phi_n+d\phi_sT^s_{mn}.
\end{equation}
In addition to these equations, the vanishing of the gaugino
variation requires the 2-form $F^A$ to be of instanton type
\cite{CDev,Str,HS,DT,RC,GMW,DS}. A $G_2$-instanton in dimension seven
is a $G_2$-connection $A$ with curvature $F^A\in \frak{g}_2$. The
latter can be expressed in any of the next two equivalent ways
\begin{equation}\label{7inst}
F^A_{mn}\Theta^{mn}\hspace{0mm}_p=0 \quad \Leftrightarrow \quad
F^A_{mn}= -\frac{1}{2}F^A_{pq}(*_7\Theta)^{pq}\hspace{0mm}_{mn};
\end{equation}

\subsubsection{Theorem~\ref{main} in dimension 7}

\begin{proof}
We have to investigate the Einstein equation of motion in dimension 7.
First we show that
\begin{equation}\label{cl1}
dT_{mjkl}*_7\Theta_{njkl}=dT_{njkl}*_7\Theta_{mjkl}.
\end{equation}
Indeed, the second identity in \eqref{mo} and \eqref{tsol7} yield
\begin{equation}\label{smo7}Ric^+_{mn}-Ric^+_{nm}=(*_7d*_7T)_{mn}=-2(*_7(d\phi\wedge d\Theta))_{mn}=2(*_7(d\phi\wedge *_7T))_{mn}=2d\phi_sT^s_{mn}\end{equation}
which compared with the skew-symmetric part of \eqref{ric77} gives \eqref{cl1}. In particular, \eqref{smo7} gives a proof of the second equality in \eqref{mot} in dimension seven.

Insert \eqref{ric77} into the first equality in \eqref{mo} and use \eqref{cl1} to get
\begin{gather}\label{mo7}
Ric^g_{ij}=-2\nabla^g_id\phi_j-\frac1{12}dT_{mjkl}*_7\Theta_{njkl}+\frac14T_{mpq}T_n^{pq}.
\end{gather}
Substitute \eqref{acgen}  into \eqref{mo7}, use  \eqref{7inst} and
compare the result with the first equation in \eqref{mot} to
conclude that the supersymmetry equations \eqref{sup1} together
with the anomaly cancellation \eqref{acgen} imply the first
equation in \eqref{mot} if and only if the next equality holds
\cite{FIUV1}
\begin{equation}\label{supmot7}
R_{mstr}R_n^{str}=-\frac16\Big[R_{msij}R_{trij}+R_{mtij}R_{rsij}+R_{mrij}R_{stij}\Big]*_7\Theta_{nstr}.
\end{equation}
The 21 dimensional space of two forms $\Lambda^2(\mathbb R^7)$
decomposes into two parts, a seven dimensional part $\Lambda^2_7$
and a fourteen dimensional part $\Lambda^2_{14}$,
$\Lambda^2(\mathbb R^7)=\Lambda^2_7\oplus\Lambda^2_{14}$. The Lie
algebra $\frak{g}_2$ of $G_2$ is isomorphic to the
two-forms satisfying 7 linear equations, namely $\frak{g}_2\cong
\Lambda_{14}^2({\mathbb R}^7) =\{\beta\in \Lambda^2({\mathbb R}^7)
\vert *_7(\beta\wedge\Theta) =- \beta\}$. The space
$\Lambda^2_{14}(\mathbb{R}^7)$ can also be described as the
subspace of 2-forms $\beta$ which annihilate $*_7\Theta$, i.e.
$\beta\wedge*_7\Theta=0$.

For the curvature 2-form $R$ we have the orthogonal splitting
$R=R_7\oplus R_{14}$, where
\begin{equation}\label{splitg2}
\begin{aligned}
(R_7)_{ij}=\frac16(2R_{ij}+R_{kl}*_7\Theta_{ijkl});\qquad
(R_{14})_{ij}=\frac16(4R_{ij}-R_{kl}*_7\Theta_{ijkl}).
\end{aligned}
\end{equation}
The equality \eqref{g2-4-form} and \eqref{splitg2} imply
\begin{equation}\label{g2novo}
(R_7)_{kl}*_7\Theta_{klij} = 4(R_7)_{ij},\qquad
(R_{14})_{kl}*_7\Theta_{klij} = -2(R_{14})_{ij}.
\end{equation}
Using \eqref{g2novo}, we get from \eqref{supmot7} that
\begin{equation}\label{g2fin}
6(||R_7||^2+|R_{14}||^2)=6R_{mstr}R_m^{str}=-12||R_7||^2+6||R_{14}||^2.
\end{equation}
Consequently, \eqref{g2fin} yields $||R_7||^2=0$. Compare with the first equality in \eqref{splitg2} to conclude that $R_7=0$ is equivalent to the $G_2$-instanton condition (the second equality in \eqref{7inst}), i.e $R$ is a $G_2$-instanton.
\end{proof}

\subsection{Dimension eight. Proof of Theorem~\ref{main} in $d=8$.}
The existence of $\nabla^+$-parallel spinor in dimension 8
determines a $Spin(7)$ structure whose properties as
well as solutions to gravitino and dilatino Killing-spinor equations
are investigated in \cite{I1,GKMW,GMW,FIUV1}.

We briefly recall the notion of a $Spin(7)$ structure. Consider
${\mathbb R}^8$ endowed with an orientation and its standard inner
product. Consider the 4-form $\Phi$ on ${\mathbb R}^8$ given by
\begin{eqnarray}\label{s1}
\Phi &=&e_{0127} - e_{0236} + e_{0347}+e_{0567} - e_{0146} - e_{0245} +
  e_{0135}
 \\ \nonumber &+&
e_{3456} + e_{1457} + e_{1256}+e_{1234} + e_{2357} +
  e_{1367} - e_{2467}.\nonumber
\end{eqnarray}
The 4-form  $\Phi$ is self-dual  and the 8-form
$\Phi\wedge\Phi$ coincides with the volume form of ${\mathbb R}^8$.
The subgroup of $GL(8,\mathbb{R})$ which fixes $\Phi$ is isomorphic
to the double covering $Spin(7)$ of $SO(7)$.
 The 4-form $\Phi$ corresponds to a real spinor  and therefore, $Spin(7)$
can be identified as the isotropy group of a non-trivial real
spinor.

We have the well known formula (see e.g. \cite{GMW})
\begin{equation}\label{spin7-4-form}
\Phi_{ijpq}\Phi_{klpq} = 6\delta_{ik}\delta_{jl} -
  6\delta_{il}\delta_{jk} + 4\Phi_{ijkl}.
\end{equation}

A \emph{$Spin(7)$-structure} on an 8-manifold $M$ is by definition
a reduction of the structure group of the tangent bundle to
$Spin(7)$. This can be described geometrically by saying that
there exists a nowhere vanishing global differential 4-form~$\Phi$
on $M$ which can be locally written as (\ref{s1}).

If $\nabla^g\Phi=0$ then the holonomy of the metric
$Hol(g)$ is a subgroup of $Spin(7)$ and
$Hol(g)\subset Spin(7)$ if and only if $d\Phi=0$ \cite{F} (see
also \cite{Br,Sal}). The Lee form $\theta^8$ is defined by \cite{C1}
$
\theta^8
=-\frac{1}{7}*_8(*_8d\Phi\wedge\Phi)=\frac{1}{7}*_8(\delta\Phi\wedge
\Phi).
$

It is shown in \cite{I1} that the gravitino Killing spinor
equation always has a solution in dimension 8, i.e. any
$Spin(7)$-structure admits a unique $Spin(7)$-connection with
totally skew-symmetric torsion $T=*_8
d\Phi-\frac76*_8(\theta^8\wedge\Phi).$ Applying \cite[Corollary~6.18]{I1} and the identity
$*_8(\theta^8\wedge\Phi)=(\theta^8\lrcorner \Phi)$ we can also write
\begin{equation}\label{spin7liad}
\theta^8_s=\frac{1}{42}((*_8d\Phi)_{ijk}\Phi_{sijk})=-\frac{1}{42}(\delta\Phi_{ijk}\Phi_{sijk}),
\qquad T_{ijk}\Phi_{sijk}=-7\theta^8_s.
\end{equation}

The necessary conditions to have a solution to the system of
dilatino and gravitino Killing spinor equations were derived in
\cite{GKMW,I1}, and the sufficiency was proved in \cite{I1}. The
general  result \cite{I1} states that there exists a
non-trivial solution to both dilatino and gravitino Killing spinor
equations in dimension~8 if and only if there exists a
$Spin(7)$-structure $(\Phi,g)$ with an exact Lee form which is
equivalent to the statement that the conformal $Spin(7)$-structure
$(\bar\Phi=e^{-\frac{12}7\phi}\Phi,\bar g=e^{-\frac67\phi}g)$ has zero Lee form, $\bar\theta^8=0$.

The torsion 3-form (the flux~$H$) and the Lee form of a solution to the gravitino and dilatino equations in dimension eight are given by \cite{GKMW,I1}
\begin{equation}\label{tsol8}
H=T=*_8d\Phi- 2*_8(d\phi\wedge\Phi), \qquad \theta^8=\frac{12}{7}d\phi.
\end{equation}
The Ricci tensor of the torsion connection is calculated in
\cite{I1} (see also \cite{FIUV1})
\begin{equation}\label{ric8}
Ric^+_{mn}=\frac1{12}dT_{mjkl}\Phi_{njkl}+\frac16\nabla^+_mT_{jkl}\Phi_{njkl}.
\end{equation}
Using the special expression of the torsion \eqref{tsol8} and
\eqref{spin7liad}, the equation \eqref{ric8} takes
the form
\begin{equation}\label{ric88}
Ric^+_{mn}=\frac1{12}dT_{mjkl}\Phi_{njkl}-2\nabla^+_md\phi_n=\frac1{12}dT_{mjkl}
\Phi_{njkl}-2\nabla^g_md\phi_n+d\phi_sT^s_{mn}.
\end{equation}

In addition to these equations, the vanishing of the gaugino
variation requires the 2-form $F^A$ to be of instanton type
\cite{CDev,Str,HS,DT,RC,GMW,DS}. A $Spin(7)$-instanton in dimension
eith  is a $Spin(7)$-connection $A$ with curvature 2-form~$F^A\in
\frak{spin}(7)$. The latter is equivalent to
\begin{equation}\label{8inst}
F^A_{mn}=-\frac{1}{2}F^A_{pq}\Phi^{pq}\hspace{0mm}_{mn}.
\end{equation}

\subsubsection{Theorem~\ref{main} in dimension 8}

\begin{proof}
It is sufficient  to investigate only the Einstein equation of motion.
First we show that
\begin{equation}\label{cl2}
dT_{mjkl}\Phi_{njkl}=dT_{njkl}\Phi_{mjkl}.
\end{equation}
Indeed, the second identity in \eqref{mo} and \eqref{tsol8} yield
\begin{equation}\label{smo8}Ric^+_{mn}-Ric^+_{nm}=(*_8d*_8T)_{mn}=2(*_8(d\phi\wedge d\Theta))_{mn}=2(*_8(d\phi\wedge *_8T))_{mn}=2d\phi_sT^s_{mn}\end{equation}
which compared with the skew-symmetric part of \eqref{ric88} gives \eqref{cl2}. In particular, \eqref{smo8} supplies  a proof of the second equality in \eqref{mot} in dimension eight.

Substitute \eqref{ric88} into \eqref{mo} and use \eqref{cl2} to get
\begin{gather}\label{mo8}
Ric^g_{ij}=-2\nabla^g_id\phi_j-\frac1{12}dT_{mjkl}\Phi_{njkl}+\frac14T_{mpq}T_n^{pq}.
\end{gather}
Insert \eqref{acgen}  into \eqref{mo8}, use  \eqref{8inst} and
compare the result with the first equation in \eqref{mot} to
conclude that the supersymmetry equations \eqref{sup1} together
with the anomaly cancellation \eqref{acgen} imply the first
equation in \eqref{mot} in dimension eight if and only if the next equality holds
\cite{FIUV1}
\begin{equation}\label{supmot8}
R_{mstr}R_n^{str}=-\frac16\Big[R_{msij}R_{trij}+R_{mtij}R_{rsij}+R_{mrij}R_{stij}\Big]\Phi_{nstr}.
\end{equation}
The 28 dimensional space of two forms $\Lambda^2(\mathbb R^8)$
decomposes into two parts, a seven dimensional part $\Lambda^2_7$
and a twenty one dimensional part $\Lambda^2_{21}$,
$\Lambda^2(\mathbb R^8)=\Lambda^2_7\oplus\Lambda^2_{21}$. The Lie
algebra $\frak{spin}(7)$ of $Spin(7)$ is isomorphic to
the two-forms satisfying 7 linear equations, namely
$\frak{spin}(7)\cong \{\beta \in
\Lambda^2(\mathbb{R}^8)|*_8(\beta\wedge\Phi)=-\beta\}$.

For the curvature 2-form $R$ we have the splitting
$R=R_7\oplus R_{21}$, where
\begin{equation}\label{splitspin7}
\begin{aligned}
(R_7)_{ij}=\frac18(2R_{ij}+R_{kl}\Phi_{ijkl});\qquad
(R_{21})_{ij}=\frac18(6R_{ij}-R_{kl}\Phi_{ijkl}).
\end{aligned}
\end{equation}
The equality \eqref{spin7-4-form} and \eqref{splitspin7} imply
\begin{equation}\label{spin7novo}
(R_7)_{kl}\Phi_{klij} = 6(R_7)_{ij},\qquad
(R_{21})_{kl}\Phi_{klij} = -2(R_{14})_{ij}.
\end{equation}
Using \eqref{spin7novo}, we get from \eqref{supmot8} that
\begin{equation}\label{spin7fin}
6(||R_7||^2+||R_{14}||^2)=6R_{mstr}R_m^{str}=-18||R_7||^2+6||R_{14}||^2.
\end{equation}
Consequently, \eqref{spin7fin} yields $||R_7||^2=0$. Compare with the first equality in \eqref{splitspin7} to conclude that $R_7=0$ is equivalent the $Spin(7)$-instanton condition  \eqref{8inst}, i.e. $R$ is a $Spin(7)$-inatanton.
\end{proof}

\medskip
\noindent {\bf Acknowledgments.} I would like to thank the referee for valuable comments and remarks.
I would also like to thank Tony Pantev for many interesting discussions and Guangyu Guo for his useful comments.
I acknowledge with  thanks the hospitality and support of the Department of Mathematics at UPenn where the final stage of the research was done. This work has been partially supported through Contract 082/2009 with the University
of Sofia `St.Kl.Ohridski', Contract ``Idei", DO 02-257/18.12.2008 and ID 02-77.


\begin{thebibliography}{33}

\bibitem{Berg} E. A. Bergshoeff, M. de Roo, {\em The quartic
effective action of the heterotic string and supersymmetry}, Nucl.
Phys. {\bf B 328} (1989), 439.

\bibitem{HT} C.M. Hull, P.K. Townsend, {\em The two loop beta
function for sigma models with torsion}, Phys. Lett. {\bf B 191}
(1987), 115.

\bibitem{GPap} J. Gillard, G. Papadopoulos, D. Tsimpis, {\em
Anomaly, Fluxes and (2,0) Heterotic-String Compactifications},
JHEP 0306 (2003) 035.

\bibitem {Str} A. Strominger, {\em Superstrings with torsion}, Nucl. Phys.
{\bf B 274} (1986) 253.

\bibitem{CDev} E. Corrigan, C. Devchand, D.B. Fairlie, J. Nuyts, {\em First-order
equations for gauge fields in spaces of dimension greater than
four}, Nuclear Phys. B {\bf 214} (1983),  no. 3, 452--464.

\bibitem {HS} J.A. Harvey, A. Strominger, {\em Octonionic superstring solitons},
Phys. Review Let. {\bf 66} 5 (1991) 549.

\bibitem{DT} S.K. Donaldson, R.P. Thomas, {\em Gauge theory in higher dimensions},
 The geometric universe (Oxford, 1996), 31--47, Oxford Univ. Press, Oxford, 1998.

\bibitem{RC} R.  Reyes Carri\'{o}n, {\em A generalization of the notion of instanton},
Diff. Geom. Appl. {\bf 8} (1998), no. 1, 1--20.

\bibitem {GMW} J. Gauntlett, D. Martelli, D. Waldram, {\em Superstrings with
  Intrinsic torsion}, Phys. Rev. {\bf D69} (2004) 086002.

\bibitem{DS} S. Donaldson, E. Segal, {\em Gauge Theory in higher dimensions, II} arXiv:0902.3239 [math.DG]


\bibitem{HuW} C.M. Hull, E. Witten, {\em Supersymmetric sigma
models and the Heterotic String}, Phys. Lett. {\bf B 160} (1985),
398.

\bibitem{HP1} P.S. Howe, G. Papadopoulos, {\em Ultraviolet
behavior of two-dimensional supersymmetric non-linear sigma
models}, Nucl. Phys. {\bf B 289} (1987), 264.

\bibitem {GKMW} J. Gauntlett, N. Kim, D. Martelli, D. Waldram, {\em Fivebranes
  wrapped on SLAG three-cycles and related geometry}, JHEP 0111 (2001) 018.

\bibitem{CCDLMZ} G.L. Cardoso, G. Curio, G. Dall'Agata, D. Lust, P. Manousselis, G. Zoupanos,
{\em Non-K\"aeler string back-grounds and their five torsion
classes}, Nuclear Phys. {\bf B 652} (2003), 5--34.

\bibitem {GMPW} J.P. Gauntlett, D. Martelli, S. Pakis, D. Waldram,
{\em  G-Structures and Wrapped NS5-Branes}, Commun. Math. Phys.
{\bf 247} (2004), 421-445.

\bibitem{Car1}  G. L. Cardoso, G. Curio, G. Dall'Agata, D. Lust, {\em
BPS Action and Superpotential for Heterotic String
Compactifications with Fluxes}, JHEP 0310 (2003) 004.

\bibitem {BBDG} K. Becker, M. Becker, K. Dasgupta, P.S. Green,
{\em Compactifications of Heterotic Theory on Non-K\"ahler Complex
Manifolds: I}, JHEP 0304 (2003) 007.

\bibitem{BBE}  K. Becker, M. Becker, K. Dasgupta, P.S. Green, E. Sharpe,
{\em Compactifications of Heterotic Strings on Non-K\"ahler
Complex Manifolds: II}, Nucl. Phys. {\bf B678} (2004), 19-100.

\bibitem{BBDP} K. Becker, M. Becker, K. Dasgupta, S. Prokushkin,
{\em Properties from heterotic vacua from superpotentials},
hep-th/0304001.

\bibitem{y1} J. Li, S-T. Yau, {\em The Existence of Supersymmetric String Theory with Torsion},
 J. Diff. Geom. {\bf 70}, no. 1, (2005).

\bibitem{y2} J-X. Fu, S-T. Yau, {\em Existence of Supersymmetric Hermitian Metrics
with Torsion on Non-Kaehler Manifolds}, arXiv:hep-th/0509028.

\bibitem{y3} J-X. Fu, S-T. Yau, {\em The theory of superstring with flux on non-K\"ahler
manifolds and the complex Monge-Amp\`ere equation}, J. Diff. Geom.
{\bf 78} (2008), 369-428.

\bibitem{y4} K. Becker, M. Becker, J-X. Fu, L-S. Tseng, S-T. Yau,
{\em Anomaly Cancellation and Smooth Non-Kahler Solutions in
Heterotic String Theory}, Nucl. Phys. {\bf B751} (2006) 108-128.

\bibitem{Hull} C.M. Hull, {\em Anomalies, ambiquities and
superstrings}, Phys. lett. {\bf B167} (1986), 51.

\bibitem{Sen} A. Sen, {\em (2,0) supersymmetry and space-time
supersymmetry in the heterotic strin theory}, Nucl. Phys. {\bf B
167} (1986), 289.


\bibitem{II} P. Ivanov, S. Ivanov, {\em SU(3)-instantons and
$G_2,Spin(7)$-Heterotic string solitons}, Comm. Math. Phys. {\bf
259} (2005), 79-102.


\bibitem{Bwit} B. de Wit, D.J. Smit, N.D. Hari Dass, {\em Residual Supersymmetry
Of Compactified D=10 Supergravity}, Nucl. Phys. {\bf B 283}
(1987), 165.

\bibitem{FGW} D.Z. Freedman, G.W. Gibbons, P.C. West, {\em Ten Into Four Won't Go},
Phys. Lett. {\bf B 124} (1983), 491.

\bibitem {IP1} S. Ivanov, G. Papadopoulos, {\em Vanishing Theorems and String Backgrounds},
 Class.Quant.Grav. {\bf 18} (2001) 1089-1110.

\bibitem {IP2} S. Ivanov, G. Papadopoulos, {\em A no-go theorem for string warped compactifications},
 Phys.Lett. {\bf B497} (2001) 309-316.


\bibitem {FIUV2} M. Fern\'andez, S. Ivanov, L. Ugarte, R. Villacampa, {\em
Compact supersymmetric solutions of the heterotic equations of
motion in dimension 5}, Nuclear Physics  B {\bf 820} (2009), 483-502.


\bibitem {FIUV} M. Fern\'andez, S.Ivanov, L. Ugarte, R. Villacampa, {\em Non-Kaehler heterotic-string
compactifications with non-zero fluxes and constant dilaton},
Commun. Math. Phys. {\bf 288} (2009), 677-697; arXiv:0804.1648.

\bibitem {FIUV1} M. Fern\'andez, S.Ivanov, L. Ugarte, R. Villacampa, {\em
Compact supersymmetric solutions of the heterotic equations of
motion in dimensions 7 and 8}, arXiv:0806.4356.

\bibitem{Do} S.K. Donaldson, {\em Infinite determinants, stable bundles and curvature}, Duke Math. J. {\bf 54} (1987), 231-247

\bibitem{UY} K. Uhlenbeck, S.-T. Yau, {\em On the existence of Hermitian-Yang-Mils connections on stable vector bundles}, Comm. pure Appl. Math. {\bf 39} (1986) no. S, suppl., S257-S293.

\bibitem{LY} J. Li, S.-T. Yau, {\em Hermitian-yang-Mills connections on non-K\"ahler manifolds}, Math. Aspects of string theory (S.-T. Yau editor), World Scient. Publ. London 1987, 560-573.

\bibitem{KO} H. Kunitomo, M. Ohta, {\em Supersymmetric $AdS_3$ solutions in Heterotic Supergravity}, arXiv:0902.0655 [hep-th].


\bibitem{BBCG} K. Becker, C. Bertinato, Y.-C. Chung, G. Guo, {\em Supersymmetry breaking, heterotic strings and fluxes}, arXiv:0904.2932 [hep-th].

\bibitem{GuP} J. Gutowski, G. Papadopoulos, {\em Heterotic black horizons}, arXiv:0912.3742 [hep-th].

\bibitem {FI} Th. Friedrich, S. Ivanov, {\em Parallel spinors and connections
  with skew-symmetric torsion in string theory}, Asian J. Math. {\bf 6} (2002), 3003-336.

\bibitem{FI2} Th. Friedrich, S. Ivanov, {\em Almost contact manifolds, connections with torsion,
parallel spinors}, J. reine angew. Math. {\bf 559} (2003),
217-236.


\bibitem {GIP} J. Gutowski, S. Ivanov, G. Papadopoulos,
{\em Deformations of generalized calibrations and compact non-Kahler
manifolds with vanishing first Chern class}, Asian J. Math. {\bf 7}
(2003), 39-80.

\bibitem{FI1} Th. Friedrich, S.Ivanov, {\em Killing spinor equations in
  dimension 7 and geometry of integrable $G_2$ manifolds}, J. Geom. Phys
  {\bf 48} (2003), 1-11.

\bibitem {I1} S. Ivanov, {\em Connection with torsion, parallel spinors and geometry of Spin(7) manifolds},
Math. Res. Lett.  {\bf 11} (2004), no. 2-3, 171--186.

\bibitem {GLP} U. Gran, P. Lohrmann, G. Papadopoulos, {\em The spinorial geometry
of supersymmetric heterotic string backgrounds}, JHEP 0602 (2006)
063.

\bibitem {GPRS} U. Gran, G. Papadopoulos, D. Roest, P. Sloane, {\em Geometry
of all supersymmetric type I backgrounds}, JHEP 0708 (2007) 074.

\bibitem {GPR} U. Gran, G. Papadopoulos, D. Roest, {\em Supersymmetric
heterotic string backgrounds}, Phys. Lett. B {\bf 656} (2007),
119.

\bibitem {P} U. Gran, G. Papadopoulos, {\em Solution of heterotic Killing spinor equations
and special geometry}, arXiv:0811.1539 [math.DG].


\bibitem{Bl} D. Blair, {\em Contact manifolds in Riemannian geometry},
Lect. Notes Math. {\bf 509}, Springer-Verlag, 1976.

\bibitem{CM} D. Chinea, J.C. Marrero, {\em Classifications of almost contact metric structures},
Rev. Roumaine Math. Pures Appl. {\bf 37} (1992), 581-599.

\bibitem{ConS}
D. Conti, S. Salamon, {\em Generalized Killing spinors in
dimension $5$}, Trans. Amer. Math. Soc. {\bf 359} (2007),
5319--5343.

\bibitem{GGMPR} G. Gauntlett, J. Gutowski, C. Hull, S. Pakis, H.
Reall, {\em All supersymmetric solutions of minimal supergravity
in five dimensions}, Class. Quantum Grav. {\bf 20} (2003),
4587-4634.


\bibitem{Br1} R.~L. Bryant, \emph{{Some remarks on $G_2$-structures}},
  Proceeding of G\"okova Geometry-Topology Conference 2005
  (S.~Akbulut, T.~\"Onder, and R.J. Stern, eds.), International Press,
  2006.

\bibitem{CI1} R. Cleyton, S. Ivanov,  \emph{On the geometry of closed $G_2$-structures}, Commun.
  Math. Phys. \textbf{270} (2007), no.~1, 53--67.

\bibitem{CI2} R. Cleyton, S. Ivanov,  \emph{Curvature decomposition of $G_2$ manifold}, J. Geom. Phys.
{\bf 58} (2008), 1429-1449.

\bibitem {Gr} A. Gray, {\em Vector cross product on manifolds}, Trans. Am.
  Math. Soc.  {\bf 141} (1969), 463-504, Correction {\bf 148} (1970), 625.

\bibitem{FG} M.~Fern{\'a}ndez, A.~Gray, \emph{Riemannian manifolds with structure group
  {$G\sb{2}$}}, Ann. Mat. Pura Appl. (4) \textbf{132} (1982), 19--45 (1983).


\bibitem{Br}
R.~L. Bryant, \emph{Metrics with exceptional holonomy}, Ann. of Math. (2)
  \textbf{126} (1987), no.~3, 525--576.

\bibitem{Sal} S. Salamon, Riemannian geometry and holonomy groups, Pitman
  Res.  Notes Math. Ser., 201 (1989).

\bibitem{Cabr} F. Cabrera, {\em On Riemannian manifolds with $G_2$-structure},
Bolletino UMI A {\bf 10} (7) (1996), 98-112.




\bibitem {F} M. Fernandez, {\em A classification of Riemannian manifolds with structure group
$Spin(7)$}, Ann. Mat. Pura Appl. {\bf 143} (1982), 101-122.


\bibitem {C1} F. Cabrera, {\em On Riemannian manifolds with $Spin(7)$-structure},
Publ. Math. Debrecen {\bf 46} (3-4) (1995), 271-283.

























\end{thebibliography}
\end{document}